\documentstyle[12pt]{article}
\newcommand{\newc}{\newcommand} 
\newc{\ra}{\rightarrow} 
\newc{\lra}{\leftrightarrow} 
\newc{\beq}{\begin{equation}} 
\newc{\eeq}{\end{equation}} 
\newc{\barr}{\begin{eqnarray}} 
\newc{\earr}{\end{eqnarray}} 
\newcommand{\sj}[6]{\mbox{$\left \{ \begin{array}{ccc}#1 &#2 &#3 \\
#4 &#5 &#6 \end{array} \right \}$ }}

\begin{document}
\begin{titlepage}

\title{\Large \bf The exotic double-charge exchange $\mu^-\to e^+$ conversion 
in nuclei } 

\author{P.C. Divari, J.D. Vergados, T.S. Kosmas \\
{\small\it Department of Physics, University of Ioannina}\\
{\small\it GR-45110 Ioannina, Greece} \\[0.4cm] L.D.  Skouras\\
{\small\it Institute of Nuclear Physics, NCSR Demokritos}\\
{\small\it GR-15310, Aghia Paraskevi, Greece}}
\date{\today}
\maketitle
\begin{abstract}
\noindent
The formalism for the neutrinoless ($\mu^-,e^+$) conversion 
is investigated in detail and the relevant nuclear matrix elements for
light intermediate neutrinos in the case of $^{27}Al$$(\mu^-,e^+)$$^{27}Na$ 
are calculated. The nucleus $^{27}$Al
is going to be used as a stopping target in the MECO experiment at 
Brookhaven, one of the most sensitive probes expected to reach a
sensitivity in the branching ratio  of the order $10^{-16}$ within the next
 few years.
The relevant transition operators are constructed utilizing a variety of
mechanisms present in current gauge theories, with emphasis on the intermediate
neutrinos, both light and heavy, and heavy SUSY particles. 
The nuclear wave functions, both for the initial state and all excited final
states are obtained in the framework of $1s-0d$ shell model employing the
 well-known and
tested Wildenthal realistic interaction. In the case of the light intermediate
neutrinos the transition rates to all excited final states up to $25~MeV$
in energy are calculated. We find that the imaginary part of the amplitude is 
dominant. The total rate is calculated by summing over all
 these partial transition strengths. We also find that the rate due to the
real part of the amplitude is much smaller than the corresponding quantity
found previously by the closure approximation.
\end{abstract}

\bigskip
\noindent
PACS number(s): 23.40.Bw, 23.40.-s, 14.60.Pq, 21.60.Cs  

\bigskip
\noindent
\parskip=0mm
\parindent=1.5em
\end{titlepage}

\newpage
\section{Introduction}

Modern gauge theories (grand unified theories, supersymmetry, etc.), 
which go beyond the standard model (SM), predict a great number of processes 
which violate the lepton and/or lepton family (flavor) quantum numbers 
\cite{Scheck}-\cite{KLV94}. One can distinguish three categories of such
processes: the purely leptonic ($\mu \to e +\gamma$, $\mu \to 3 e $, etc.),
the semileptonic hadron decays ($K_L^0\to \mu^\pm e^\pm$, etc.), and 
the semileptonic which take place in the presence of nuclei.
Among the most interesting such processes are those which 
take place in a muonic atom \cite{Dep}-\cite{Schaaf}. 
One exotic such process is
the muon-to-positron conversion, 
\begin{equation}
\label{1}
\mu^-  + (A,Z)  \rightarrow  e^+ + (A,Z-2)  \, ,
\end{equation}
which violates the muonic ($L_\mu$), electronic ($L_e$) and total lepton ($L$) 
quantum numbers \cite{Abe}-\cite{JDV91}. Another exotic process is the
muon-to-electron conversion, 
\begin{equation}
\label{2}
\mu^-  + (A,Z)  \rightarrow  e^- + (A,Z)  \, ,
\end{equation}
which violates only the lepton-family
quantum numbers $L_\mu$ and $L_e$ \cite{Molzon,SSK99}. 

Both of these processes can experimentally be studied simultaneously, since
 both of them have
the same intrinsic background and the same initial state (a muon at rest in the 
innermost 1S orbit of a muonic atom).
In the present work we will focus our attention on reaction  (\ref{1}). 

In recent years, continuous experimental efforts have been devoted for the measurement
of the branching ratio $R_{\mu e^+}$ defined as the ratio of the 
$(\mu^-,e^+$) conversion rate divided by the total 
rate of the ordinary muon capture reaction \cite{Blin}, i.e.
\begin{equation}
\label{rat}
R_{\mu e^+}=\Gamma(\mu^-\to e^+)/\Gamma(\mu^-\to\nu_\mu).
\end{equation}
Up to now only upper limits have been set 
and the best limit found is for the $^{48}$Ti nucleus at TRIUMF 
and PSI \cite{Bry72,Bader,Ahmad,Schaaf} yielding the values 
$$
R_{\mu e^+} \leq 4.6 \times 10^{-12} \, \, \, \, \cite{Ahmad}
$$
and  
$$
R_{\mu e^+} \leq 4.4 \times 10^{-12} \, \, \, \, \cite{Dohmen} \, ,
$$ 
respectively. This limit could be further improved by future experiments, 
at PSI (SINDRUM II experiment), which aims to push the sensitivity of the branching 
ratio $R_{\mu e^+}$ to $10^{-14}$, and at Brookhaven (MECO experiment) \cite{Molzon}
with expected sensitivity about three to four orders of magnitude 
below the existing experimental limits \cite{Molzon,SSK99}. 

Traditionally the exotic $\mu^-\to e^\pm$ processes were searched by employing
 medium heavy
(like $^{48}$Ti and $^{63}$Cu) \cite{Dohmen,Schaaf} or very heavy (like $^{208}$Pb
and $^{197}$Au) \cite{Honec,Ahmad,Schaaf} targets.
For technical reasons the MECO target has been chosen to be the light nucleus $^{27}$Al
\cite{Molzon}. The MECO experiment, which is planned to start soon
at the Alternating Gradient Synchrotron (AGS), 
is going to employ a new very intense $\mu^-$ beam and a new detector
 \cite{Molzon}.
The basic feature of this experiment is the use of a pulsed $\mu^-$ beam to
significantly reduce the prompt background from $\pi^-$ and $e^-$
contaminations. 

The best upper limit for the $\mu^-\to e^-$ conversion branching ratio
$R_{\mu e^-}$ set up to the present has been extracted at
PSI (SINDRUM II experiment) as
\begin{equation}
R_{\mu e^-}^{Au} \leq 5.0\times 10^{-13} \, ,
\quad\mbox{ for }^{197}\mbox{Au target \cite{Schaaf}} \, .
\label{Au}
\end{equation}
For the $^{48}$Ti target the determined best limit is
\begin{equation}
R_{\mu e^-}^{Ti}\leq 6.1\times 10^{-13} \, \quad\mbox{\cite{Dohmen}} \, , 
\label{Ti}
\end{equation}
while for the $^{208}$Pb target the extracted limit is
\begin{equation}
R_{\mu e^-}^{Pb}\leq 4.6\times 10^{-11} \, \quad\mbox{\cite{Honec}}.
\label{Pb}
\end{equation}

Processes (1) and (2) are very good examples of the interplay between nuclear and
particle physics in the area of physics beyond the standard model. Moreover,
the ($\mu^-,e^+$) conversion has many similarities with the neutrinoless 
double $\beta$-decay ($0\nu \beta\beta$), 
\begin{equation}
\label{3}
(A,Z)  \rightarrow  e^- + e^- + (A,Z+2) \, ,
\end{equation}
and especially with its sister electron to positron conversion
\cite{JDV86}, which violate the lepton-flavor 
($L_e$) and total lepton ($L$) quantum numbers. Both 
reactions (\ref{1}) and (\ref{3}) involve a change of charge by two
units and thus they cannot occur in the same nucleon. Both of them are
forbidden, if lepton number is absolutely conserved. One can show that,
if either of these processes  is observed, the neutrinos must be
massive Majorana particles. In spite of the many similarities, however,
these double charge exchange processes have some significant
differences which can be briefly summarized as follows:

(i) Due to the nuclear masses involved, neutrinoless double beta decay
can occur only in specific nuclear systems for which single beta decay
is absolutely forbidden, due to energy conservation, or greatly hindered
due to angular momentum mismatch. These systems, with the
possible exception of $^{48}Ca$, have complicated nuclear structure.
 Furthermore, the
neutrinoless double beta decay can lead mainly to the ground state 
and, in some cases, to few low-lying excited states of the
final nucleus. Such constraints are not imposed on process (\ref{1}), 
due to the rest energy of the disappearing muon.

(ii) From the corresponding experiments, in conjunction with appropriate
 nuclear matrix elements as input, one
may extract lepton violating parameters, which depend on flavor. Thus,
in the framework of the neutrino mixing models, the amplitudes of
neutrinoless $\beta \beta$ decay and ($\mu^-,e^+$), for leptonic
currents of the same chirality, are proportional to some combination
of the neutrino masses, which are 
different  from each other. The same is true in the
case of the mass independent lepton violating parameters $\eta$ and $\lambda$
appearing in the amplitude, 
if the leptonic currents are of opposite chirality. 
One does not know {\it a priori} which flavor combination is favored.

(iii) The long wavelength approximation does not hold in the case of 
($\mu^-,e^+$) conversion, since the momentum of the outgoing $e^+$ is quite 
high.  Thus, the effective 
two-body operator responsible for the  ($\mu^-,e^+$) conversion is strongly 
energy dependent and more complicated than the corresponding one for 
the $0\nu \beta\beta$ decay. On the other hand, one can in this case 
choose a target,
consistent with the standard experimental requirements, so that the 
nuclear structure required is the simplest possible.

(iv) Neutrinoless double beta decay has the experimental advantage 
that there are no other fast competent channels. The only other channel
is the two-neutrino double-beta decay, which, however, is also of second order 
in weak interaction and kinematically suppressed. 

In any case we view the two processes, ($\mu^-,e^+$) conversion and 
$0\nu\beta\beta$ decay, as providing useful complementary
information. As is well known in the case of neutrinos one needs the following
information: a) the mixing matrix, b) the three eigenmasses and c) the two 
independent Majorana phases. These phases are present even in a CP conserving
theory since they can take values $0$ and $\pi$. Study of CP violation in the
leptonic sector is beyond the goals of the present paper. The interested reader
is referred to the literature, e.g. see \cite{BP83}.

 The neutrino oscillation data can, in principle,
determine the mixing matrix and the two independent mass-squared differences
(e.g. $\delta m^2_{21} = m^2_2 - m^2_1$, $\delta m^2_{31} = m^2_3 - m^2_1$). 
This has been done in a number of papers, see e.g. the recent work by Haug 
{\it et al.} \cite {HFV01} and references therein. 
They cannot determine, however, the scale of the masses, e.g. the lowest
eigenvalue $m_1$ and the two relative Majorana phases. Once $m_1$ is known,
one can find  $m_2=[\delta m^2_{21}+m^2_1]^{1/2}$  
and $m_3=[\delta m^2_{31}+m^2_1]^{1/2}$. 
The masses can be chosen positive by a proper redefinition of the neutrino
fields, i.e. by absorbing the sign into the Majorana phases. If the mixing
matrix is known, the mass $m_1$ can be determined from triton decay as follows:
\beq
(m_{\nu})_{ee} \equiv m_{\nu}=| \sum_{j=1}^{3}
 U^*_{ej}U_{ej} m^2_j |^{1/2} \, ,
\label{mass.3}
\eeq 
since the Majorana phases do not appear (this experiment cannot differentiate
between Dirac and Majorana neutrinos).
In the above expressions $U$ is the mixing matrix.

 The two lepton violating processes provide two independent linear combinations 
of the masses and the Majorana phases. In fact
\beq
\langle m_{\nu}\rangle_{ee} \equiv \langle m_{\nu}\rangle =|\sum _{j=1}^{3} U_{ej}U_{ej}e^{i\lambda _j} m_j|
\label{mass.1}
\eeq
and
\beq
\langle m_{\nu}\rangle_{\mu e+}=|\sum_{j=1}^{3}
                   U^*_{\mu j}U^*_{ej}e^{-i\lambda _j}m_j| \, .
\label{mass.2}
\eeq
 It is clear from the above expressions that only the two relative CP phases are
measurable.
  
So, the three experiments together can specify all
parameters not settled by the neutrino oscillation experiments.
So, if possible, all three of them should be pursued.

Strictly speaking $(\mu^-,e^+)$ and neutrinoless double beta decay
should be treated as two-step processes by explicitly constructing
the states of the intermediate $(A,Z\pm1)$ systems. It has been found
\cite{VPF},
however, that, for neutrinoless double beta decay, since the energy
denominators are dominated by the momentum of the virtual neutrino,
closure approximation with some average energy denominator works very
well. We expect this approximation to also hold in the case of $(\mu^-,e^+)$
conversion to sufficient accuracy.
We will, therefore, replace the intermediate nuclear energies by some 
suitable average. By summing over the partial rates of all
allowed final states of the nucleus $(A,Z-2)$ we obtain the total
rate. This will then be compared 
to that obtained by invoking closure \cite{Eric}
with some appropriate mean energy $\langle E_f\rangle$ of the 
final states.

So far, theoretically the ($\mu^-,e^+$) process has been investigated  
\cite{Eric,Leo} on the exclusive reactions  
\begin{equation}
\label{6}
^{40}Ca + \mu^- \rightarrow  e^+ + ^{40}Ar(gs)
\end{equation}
\begin{equation}
\label{7}
^{58}Ni + \mu^- \rightarrow  e^+ + ^{58}Fe(gs) \, .
\end{equation}
In these studies the partial $g.s. \rightarrow g.s.$ transition rate was
calculated by performing microscopic calculations of these nuclear 
matrix elements. On the other hand, the total transition strength to all final states 
(inclusive process) was estimated along the lines of closure approximation
and ignoring $4$-body terms \cite{Eric}.

In the present article we apply the shell-model techniques in the investigation
of the ($\mu^-,e^+$) conversion reaction
\begin{equation}
\label{Al-react}
^{27}Al + \mu^- \rightarrow  e^+ + ^{27}Na \, .
\end{equation}
To this end, we construct all the needed wave functions and calculate
the rate not only to the ground state but also to 
the excited states of the final nucleus.  As a first step we limited
ourselves to the mass mechanism for light neutrinos. So, only the
multipolarities associated with the Fermi and Gamow-Teller type operators
were relevant. For practical reasons we had
to limit ourselves to  states lying below some excitation energy 
($\approx 25$ MeV) of the final nucleus $^{27}$Na. 
The distribution of the strengths for the most important
multipolarities versus the excitation energy of the daughter nucleus in the 
reaction (\ref{Al-react}) is also studied. 

The paper is organized as follows. 
In sect. 2, an extensive presentation of the relevant expressions occurring 
in the formal description of the $\mu^-\to e^+$ transition operators 
is given. In sect. 3, we deal with the expressions of the branching
ratios. In sect. 4, we discuss the evaluation of the
inclusive $\mu^-\to e^+$ matrix elements by means of explicit
construction of the needed nuclear wave functions in the framework of 
the $s-d$ shell model. 
In sect. 5, the results obtained for the transition matrix elements 
in the case of $^{27}Al (\mu^-, e^+)^{27}Na$ are presented and discussed.
Also the spreading of the contributions due to the occurrence of various 
multipoles are described. Our conclusions are summarized in sect. 6.

\section{Brief theoretical formulation of the $\mu^-\to e^+$ conversion operators}

\subsection{Effective $\mu^-\to e^+$ conversion Lagrangian in Gauge models}

>From the particle physics point of view, processes like $\mu^-\to e^\pm$ conversions  
are forbidden in the SM by total-lepton and/or lepton-flavor (muonic and electronic)
quantum number conservation. They have
been recognized long time ago as important probes for studying  the lepton and
lepton-flavor changing charged-current interactions \cite{Marci}-\cite{KLV94},
which go beyond the standard model.

There are several possible elementary particle mechanisms which can
mediate the lepton-violating process (\ref{1}). The mechanisms which
have been studied theoretically are: 

(i) Those mediated by a massive Majorana neutrino. In this case we
have two possibilities.

1) The chiralities of the two leptonic
currents are the same. Then, the  amplitude in the case of light
neutrinos is proportional to some average neutrino mass or to some 
average of the inverse of the neutrino mass, if it is heavy.

2) The chiralities of the leptonic currents are opposite. Then, the amplitude 
is not explicitly dependent on the neutrino mass. It vanishes, however, if
the neutrinos are not Majorana particles. This mechanism is
significant only in the case of light neutrinos. 

(ii) Those accompanied by massless or light physical Higgs particles (majorons). 

(iii) Those involving more exotic intermediate Higgs particles. 

(iv) Those mediated by intermediate supersymmetric (SUSY) particles. 

 From a nuclear physics point of view one has to be a bit careful when
the intermediate particles are very heavy. The elementary amplitude is 
constructed at the quark level, but the calculation is performed at the nucleon
level. If, in going from the quark
to the nucleon level, the nucleons are treated as point-like particles,
the nuclear matrix elements
are suppressed due to the presence of short range correlations.
To avoid this suppression, a cure has been proposed \cite{JDV86,JDV81} which 
treats the nucleons as composite
particles described by a suitable form factor. A different approach 
is to consider mechanisms, which involve particles
other than nucleons in the nuclear soup. Such are, e.g.,
mechanisms whereby the processes (\ref{1}) and (\ref{3}) are mediated 
by the decay of the doubly-charged $\Delta^{++}$ $(3/2,3/2)$ resonance
\cite{Eric,Shu,JDV91} present in the nuclear medium, i.e. 
\begin{equation}
\label{4}
\mu^- + \Delta^{++} \rightarrow n + e^+ 
\end{equation}
This process, however, to leading order does not contribute to
$0^+\rightarrow0^+$ $\beta \beta$ decays,  but it may
contribute to ($\mu^-,e^+$). 
The other process is induced by pions in flight between the two nucleons
 \cite{JDV86,JDV82,FKSS}
according to the following modes:

(a) The  1-pion mode represented by the reactions:
\begin{equation}
\label{1pi-mode}
\mu^- + p \rightarrow n + \pi^- + e^+ \, , \qquad
\pi^- + p \rightarrow n  
\end{equation}
where the protons and neutrons are bound in the nucleus.

(b) The 2-pion mode represented by the reactions
\begin{equation}
\label{2pi-mode}
p \rightarrow n + \pi^+ \,, \qquad
\pi^+ \mu^- \rightarrow \pi^- + e^+ \, , \qquad
\pi^- + p \rightarrow n  
\end{equation}

\subsection{The transition operators at nuclear level}

The gauge models mentioned in the previous subsection
give rise to a plethora of effective transition operators. 
Their essential isospin, spin and radial structure is given as follows.

1) The isospin structure is quite simple, i.e. of the form
$\tau_-(i) \tau_-(j)$ where i and j label the nucleons participating in the
 process. 

2) The spin structure is given in terms of the operators

\begin{equation}
\label{13}
W_{S1}(ij)=1 \quad  (Fermi)
\end{equation}
\begin{equation}
\label{14}
W_{S2}(ij)={\bf {\sigma}}_i \cdot {\bf \sigma}_j \quad (Gamow-Teller)
\end{equation}
\begin{equation}
\label{15}
W_{S3}(ij)=3({\bf \sigma}_i\cdot \hat{r}_{ij})({\bf \sigma}_j
\cdot \hat{r}_{ij})-{\bf \sigma}_i
\cdot {\bf \sigma}_j \quad (Tensor)
\end{equation}
\begin{equation}
\label{17}
W_{A1}(ij)=\imath {\bf \sigma}_i\times {\bf \sigma}_j
\end{equation}
\begin{equation}
\label{18}
W_{A2}(ij)={\bf \sigma}_i- {\bf \sigma}_j 
\end{equation}
The operator $\hat{r}_{ij}$ is determined below.

3) The orbital part can be expressed in terms of the quantities:

(a) The momentum of the emitted positron ($p_e$) obtained from
the kinematics of the reaction (\ref{1}). One finds that 
\begin{equation}
\label{11}
p_e \equiv \vert {\bf p}_e \vert = m_{\mu} - \epsilon_b + Q - E_x
\end{equation}
where $Q=M(A,Z)-M(A,Z-2)$ is the atomic mass difference between the initial, 
$(A,Z)$ and final, $(A,Z-2)$ 
nucleus, $\epsilon_b$ is the binding energy of the muon at the muonic atom 
($\epsilon_b \approx 0.5$ MeV for $^{27}$Al), $E_x$ is the excitation energy 
($E_x=E_f -E_{gs}$) of the final nucleus and $m_\mu$ is the muon mass 
($m_\mu = 105.6$ MeV). 

(b) The relative (${\bf r}_{i j}$) and center of mass (${\bf R}_{i j}$)
coordinates which are written as 

$${\bf r}_{ij}={\bf r}_i -{\bf r}_j \, , \qquad
{\hat r}_{ij}= \frac{ {\bf r}_{i j}} {\vert {\bf r}_{i j} \vert} \, , \qquad
 r_{i j} = \vert {\bf r}_{i j} \vert \, ,
$$ 
and 
$${\bf R}_{ij}=\frac{1}{2}({\bf r}_i +{\bf r}_j) \, , \qquad
{\hat R}_{ij}= \frac{ {\bf R}_{i j}} {\vert {\bf R}_{i j} \vert} \, , \qquad
 R_{i j} = \vert {\bf R}_{i j} \vert \, ,
$$

The radial part of the operator contains:

 (i) the spherical Bessel functions
$j_l({p_er_{ij}}/2)$ and $j_{\cal L}(p_eR_{ij})$ resulting from
the decomposition of the outgoing positron and

 (ii) a function $f(r)$ of the relative coordinate ($r=r_{ij}$) given by:
\begin{equation}
\label{20a}
f(r)=\frac{R_0}{r}F(r)\Psi_{cor}(r), 
\end{equation}
where the constant $R_0$ represents the nuclear radius ($R_0 = 1.2 A^{1/3}$).
The function $\Psi_{cor}(r)$ is some reasonable two-nucleon correlation 
function, e.g. of the type \cite{JDV86} 
\begin{equation}
\label{21}
\Psi_{cor}(r) = 1- e^{-ar^2}(1-br^2)
\end{equation}
with $ a=1.1 fm^{-2}$ and $b=0.68 fm^{-2}$. 

We mention that, strictly speaking in the above radial function the muon wave
 function must also be included.
In fact, if muon is considered as relativistic particle an additional
lepton-spin dependence appears in the transition operator.
In the case of the light nucleus $^{27}$Al studied in the present work,
 however, the muon is in the $1s$ atomic orbit. Its wave function
varies very slowly inside the nucleus and thus it can be replaced by its
 average value. Anyway this average value drops out 
if the same approximation is assumed for the $\mu^-$-capture, i.e. for the 
denominator of the branching ratio $R_{\mu e^+}$.

As we have already indicated, the radial function $F(r)$ depends on the 
specific mechanism assumed for the
$\mu^- \to e^+$ conversion process to occur. The following cases are 
of interest.

(i) In the case of light Majorana neutrinos, when the leptonic currents 
are left-handed, $F(r)$ takes the form \cite{DIVA00,SIMKO00}
\begin{equation}
\label{new-Fr}
F(r)= \frac{2}{\pi} \int_0^{\infty}\frac{sinx}{x-\alpha +\imath \epsilon} dx +
      \frac{2}{\pi} \int_0^{\infty}\frac{sinx}{x+\delta}_e dx 
\end{equation}
The quantities $\delta_e$ and $\alpha$ are given in terms of the nuclear 
masses and the average excitation energy of the intermediate states,
$\langle E_{xn}\rangle$, as
$$
\delta_e=[\langle E_{xn}\rangle +M(A,Z-1)-M(A,Z)+p_e]r
$$ 
$$
\alpha=[m_{\mu}+M(A,Z)-M(A,Z-1)-\langle E_{xn}\rangle ]r
$$
Note that $\delta_e$ depends on the positron momentum. The first term of 
$F(r)$ in Eq. (\ref{new-Fr}) can be written as
\begin{equation}
\label{23}
\frac{2}{\pi}\int_0^{\infty}\frac{sinx}{x-\alpha +\imath \epsilon} dx =
\frac{2}{\pi}~P~\int_0^{\infty}\frac{sinx}{x-\alpha} dx - 
{\imath} 2 sin{\alpha} 
\end{equation}
As can be seen, the amplitude has now an imaginary part, a fact 
that was missed in earlier calculations \cite{Leo}.
The principal value integral can be written in an equivalent form 
as follows:
\begin{equation}
\label{24}
\frac{2}{\pi}~P~\int_0^{\infty}\frac{sinx}{x-\alpha} dx  
=2cos{\alpha} -1 + \frac{2}{\pi}\alpha
\int_0^{\infty}\frac{sinx}{x(x+\alpha)} dx 
\end{equation}
The latter expression is more convenient for numerical integration techniques. 
By combining Eqs. (\ref{new-Fr})-(\ref{24}) we can write $F(r)$ with
separate real and imaginary parts as
\begin{equation}
F(r) = 2cos{\alpha} -1 + \frac{2}{\pi} \int_0^{\infty} \ {sinx} \ 
\Big[ \frac{\alpha}{x(x+\alpha)} + \frac{1}{x+\delta}_e \Big] dx \
- \ {\imath} 2 sin{\alpha} \, .
\label{Fr-Re-Im}
\end{equation}
It is worth remarking that, in the case of 
$0\nu \beta\beta$ decay $\alpha \sim 0$ therefore $F(r)=1$. 
This simplifies quite well the calculations in the 
$0\nu \beta\beta$ decay process.

(ii) In the case of light Majorana neutrinos, when the leptonic 
currents are of opposite chirality we have
$F(r)\rightarrow F^{'}(r)=r(d/dr)F(r)$. 
The same situation occurs in the context of R-parity violating
 supersymmetric interactions mediated by light Majorana neutrinos
in addition to other SUSY particles.

(iii) For heavy intermediate particles, e.g. heavy Majorana neutrinos, 
we will examine two modes:

1) Only nucleons are present in the nucleus. Then the function
 $F(r)$ reads 
\begin{equation}
\label{25}
F(r)=\frac{1}{48}\frac{m_A^2}{m_e m_p}x_A(x_A^2+3x_A+3)e^{-x_A},
\qquad x_A=m_Ar
\end{equation}
with $m_e$, $m_p$ the
masses of electron and proton respectively. It should be mentioned that
the above expression was obtained for neutrino masses much heavier than the
proton mass provided that the nucleon is not point like. The particular
 expression holds, if
the nucleon is assumed to have a finite size adequately described \cite{JDV83} 
by a dipole shape form factor with characteristic mass $m_A$ taking the 
value $m_A=0.85 GeV/c^2$

2) The process is mediated by pions in flight between the two
 interacting nucleons. Then one distinguishes two possibilities
\cite{JDV86,Vergados00}: 

(a) The  1-pion mode Eq. (\ref{1pi-mode}).
In this mode $F(r)$ is replaced by $F_{1\pi}^i$, $i=GT$ (Gamow-Teller), 
$T$ (Tensor) where
\begin{equation}
\label{25a}
F^{GT}_{1 \pi}(x)=\alpha_{1 \pi}~e^{-x}~~,~~
F^{T}_{1 \pi}(x)=\alpha_{1 \pi}~(x^2+3x+3)e^{-x}/x^2
\end{equation}
with 
$x \equiv x_{\pi} = m_{\pi}r$ ($m_\pi$ denotes the pion mass)
and $\alpha_{1 \pi}=1.4\times 10^{-2}$. In this
case the radial functions are the same with those entering the
neutrinoless double beta decay. 

(b) The 2-pion mode Eq. (\ref{2pi-mode}).
Now the radial functions are very different from those appearing in $0\nu$
$\beta \beta$ decay, since the momentum carried away by the outgoing lepton
is not negligible. It is, however, obtained from those 
entering the neutrinoless double beta decay, via the substitution:
\begin{equation}
\label{25b}
F(r)j_l(x_e /2)\rightarrow 
\int_0^1 j_l((\xi-1/2)x_e)F^i_{2\pi}([\xi(1-\xi)x^2_e+x^2_{\pi}] 
^{1/2})d\xi
\end{equation}
($ x_{e} = m_{e}r$)
where $F^i_{2 \pi},~i=GT,T$ are given by \cite{JDV86}  
\begin{equation}
\label{25c}
F^{GT}_{2 \pi}(x)=\alpha_{2 \pi}(x-2)e^{-x}~~,~~
F^{T}_{2 \pi}(x)=\alpha_{2 \pi}(x+1)e^{-x}
\end{equation}
with $\alpha_{2 \pi}=2.0\times 10^{-2}$.

\subsection{Irreducible Tensor Operators}

In this section we are going to exhibit the structure of the various 
irreducible tensor operators relevant to our calculations. We characterize 
them by the set of quantum numbers $l,{\cal L},\lambda,\mu,\Lambda,L,S,J$, some 
of which may be redundant in some special cases. Thus $l$ and ${\cal L}$ refer
 to the multipolarity of the outgoing lepton in the relative and center of mass
systems. $\lambda$ is the orbital rank of the operator in the relative
 coordinates (the corresponding rank in the CM system is ${\mu}$). $L$ is the
total orbital rank, $S$ is the spin rank and $J$ the total angular momentum
 rank. Finally $\Lambda$ is the rank of the spherical harmonic describing the
momentum of the outgoing lepton. The latter couples to zero or 1 with
the J-rank, i.e. ($\Lambda,J)k$, $k=0,1$. 
Some details on how 
these operators are combined to give the nuclear matrix elements will 
be discussed in the Appendix. 

We will begin with operators appearing when the chiralities of
the two leptonic currents involved are the same. This covers the case
of minimal left-handed extensions of the standard model.

One encounters Fermi-type operators, $\Omega_{F}$, of the form
\begin{eqnarray}
\label{29}
\Omega_F= & \sum_{i<j}\tau_-(i) \tau_-(j)
f(r_{ij})j_l(\frac{p_er_{ij}}{2})j_{\cal L}(p_eR_{ij}) 
\nonumber \\
& \left [ \sqrt{4 \pi}Y^{\lambda}(\hat{r}_{ij})\otimes \sqrt{4 \pi} Y^{\mu}(\hat{R}_{ij})
\right ]^J, 
\nonumber \\
& \quad \lambda=l, \mu={\cal L}, \quad S=0,\quad J=L
\end{eqnarray}
($\Lambda$ is redundant).
The Gamow-Teller operators, $\Omega_{GT}$, are  similarly written as 
\begin{eqnarray}
\label{29a}
\Omega_{GT}= &\sum_{i<j}\tau_-(i) \tau_-(j)
f(r_{ij})j_l(\frac{p_er_{ij}}{2})j_{\cal L}(p_eR_{ij}) 
\nonumber \\
& \left [ \left [ \sqrt{4 \pi}Y^{\lambda}(\hat{r}_{ij})\otimes \sqrt{4 \pi}
 Y^{\mu}(\hat{R}_{ij})
\right ]^L \otimes  (-\sqrt{3})\left [ {\bf \sigma}_i \otimes
{\bf \sigma}_j \right ]^0 \right ]^J,
\nonumber \\
& ~~\lambda=l,  \mu={\cal L}, \quad S=0,\quad J=L
\end{eqnarray}
($\Lambda$ is redundant).
Note that 
\begin{equation}
\label{29b}
{\bf {\sigma}}_i \cdot {\bf \sigma}_j=-\sqrt{3}\left [ {\bf \sigma}_i \otimes
{\bf \sigma}_j \right ]^0_0
\end{equation}
 The first spin antisymmetric operator is
\begin{eqnarray}
\label{30.b}
\Omega_{A1}=& \sum_{i<j}\tau_-(i) \tau_-(j)
f(r_{ij})j_l(\frac{p_er_{ij}}{2})j_{\cal L}(p_eR_{ij}) 
\nonumber \\
& \left [ \left [ \sqrt{4 \pi}Y^{\lambda}(\hat{r}_{ij})\otimes \sqrt{4 \pi} 
Y^{\mu}(\hat{R}_{ij})
\right ]^L \otimes  (-\sqrt{2})\left [ {\bf \sigma}_i \otimes
{\bf \sigma}_j \right ]^1 \right ]^J, 
\nonumber \\
& ~~\lambda=l, \mu={\cal L},
\quad S=1,\quad  J=L,|L\pm 1|
\end{eqnarray}
($\Lambda$ is redundant).
Note that
\begin{equation}
\label{30.c}
\imath{\bf \sigma}_i \times {\bf \sigma}_j=
(-\sqrt{2})\left [ {\bf \sigma}_i \otimes
{\bf \sigma}_j \right ]^1
\end{equation}
 The second spin antisymmetric operator is
\begin{eqnarray}
\label{30.d}
\Omega_{A2}=& \sum_{i<j}\tau_-(i) \tau_-(j)
f(r_{ij})j_l(\frac{p_er_{ij}}{2})j_{\cal L}(p_eR_{ij}) 
\nonumber \\
& \left [ \left [   \sqrt{4 \pi}Y^{\lambda}(\hat{r}_{ij})\otimes \sqrt{4 \pi}
Y^{\mu}(\hat{R}_{ij}) \right ]^L 
\otimes \left ( {\bf \sigma}_i - {\bf \sigma}_j \right ) \right ]^J,
\nonumber \\ 
& ~~\lambda=l, \mu={\cal L}, \quad S=1,\quad  J=L,|L\pm 1|
\end{eqnarray}
($\Lambda$ is redundant).
Note that each operator must be overall symmetric with respect to
interchange of the particle indices. So, in those cases
in which the spin operator is of rank unity, $l$ must be odd. 
In the special case of $0^+\rightarrow0^+$ neutrinoless double beta decay, 
only the Fermi and Gamow-Teller operators occur.

We are now going to consider the case in which the theory contains
both $R$ (Right) and $L$ (Left) currents. If both currents are right-handed
the above results hold, but the relevant neutrinos are heavy. We thus need only
consider the case in which we have 
$L-R$ interference in the leptonic sector. This may be important in
the case of light neutrinos. 
As we have already mentioned, this also occurs in the context of R-parity 
violating supersymmetric interactions, which,
in addition to other SUSY particles, involve intermediate
light Majorana neutrinos.
The amplitude now is proportional to
the 4-momentum of the intermediate neutrino. The time component
has a structure similar to the above with a different energy dependence.
The corresponding operators are indicated by putting a ``{\it prime}'' over the
corresponding ones for the mass term. This point will not be further pursued, 
 since their contribution is suppressed. Its space component, after the
Fourier transform, gives an amplitude proportional to the gradient
of the Fourier transform of the previous case.
We thus get the above operators, to be denoted by $\Omega^{'}_F$
$\Omega^{'}_{GT}$, and $\Omega^{'}_{A2}$ 
(associated with the term linear in the spin), with $F(r)$ replaced by
 $F^{'}(r)$.
Now the overall operator has an overall rank of a vector, obtained by the
coupling of the two operator ranks $J$ and $\Lambda$.
In this case, in addition to operators of the above form, we
encounter an operator of spin rank two, which 
is of the form:
\begin{eqnarray}
\label{30.a}
\Omega^{'}_T=& \sum_{i<j}\tau_-(i) \tau_-(j)
f^{'}(r_{ij})j_l(\frac{p_er_{ij}}{2})j_{\cal L}(p_eR_{ij}) 
\nonumber \\
 &  \left [ \left [   \sqrt{4 \pi}Y^{\lambda}(\hat{r}_{ij})\otimes \sqrt{4 \pi}
 Y^{\mu}(\hat{R}_{ij}) \right ]^{L} 
 \otimes \left [ {\bf \sigma}_i \otimes {\bf \sigma}_j \right ]^2 
\right ]^J,
\nonumber \\
 &  \, \, \, \lambda=|l \pm 1|, \mu={\cal L}, \quad S=2,\quad 
 J=L,|L\pm 1|,|L\pm 2|
\end{eqnarray}
($\Lambda$ is redundant).

As it has already been mentioned, in the case of heavy
intermediate particles one may have to consider pions in flight between
nucleons. Then one encounters only Gamow-Teller  and tensor operators
except that now the radial part is different (see Eq. (\ref{25a})-(\ref{25c})).

In the special case of $0^+\rightarrow0^+$ neutrinoless double beta 
decay mediated by light neutrinos one can invoke
the long wavelength approximation. Thus, to leading order one finds 
(up to normalization constants and possibly factors of $p_e$) 
the familiar operators:
\begin{equation}
\label{12}
\Omega_F =\sum_{i\ne j} \tau_-(i) \tau_-(j) f(r_{ij}) \quad  (Fermi)
\end{equation}
\begin{equation}
\label{14a}
\Omega_{GT} =\sum_{i\ne j} \tau_-(i) \tau_-(j) f(r_{ij})
{\bf {\sigma}}_i \cdot {\bf \sigma}_j \quad (Gamow-Teller)
\end{equation}
\begin{equation}
\label{16}
\Omega^{'}_{A2} =\sum_{i\ne j} \tau_-(i) \tau_-(j) f^{'}(r_{ij}) \,
({\bf \sigma}_i- {\bf \sigma}_j).(\imath\hat{r}_{ij}\times \hat{R}_{ij})
\end{equation}
\begin{equation}
\label{15a}
\Omega^\prime_T =\sum_{i\ne j} \tau_-(i) \tau_-(j) f^{'}(r_{ij})
\left[ 3({\bf \sigma}_i\cdot \hat{r}_{ij})({\bf \sigma}_j
\cdot \hat{r}_{ij})-{\bf \sigma}_i
\cdot {\bf \sigma}_j \right]  \quad (Tensor)
\end{equation}

\section{Branching ratio}

The branching ratio $R_{\mu e^+}$ of the ($\mu^-,e^+$) reaction
defined in Eq. (\ref{rat}) contains the LFV-parameters of the specific
gauge model assumed. These parameters, 
are entered in $R_{\mu e^+}$ via a single lepton-violating parameter $n_{eff}$. 
Under some reasonable assumptions these parameters 
can be separated from the nuclear physics aspects of the problem. As has been 
pointed out \cite{Leo}, the branching ratio $R_{\mu e^+}$ takes the form
\begin{equation}
\label{9}
R_{\mu e^+} = \rho | \eta_{eff}|^2 \; \frac{1}{A^{2/3} Z f_{PR}(A,Z)} \; 
\left(\frac{(p_e)_{max}}{m_{\mu}}\right)^2 \sum_f 
\left( \frac{p_e}{(p_e)_{max}}\right)^2 |{\cal M}_{i \rightarrow f}|^2
\end{equation}
with $ \eta_{eff}= \langle m_{\nu}\rangle_{\mu e+}/m_e$
The parameters $\rho$ and $\eta_{eff}$, which depend on the gauge model
adopted and the mechanism prevailing, are expected to be very small 
due to the fact that $\mu^-\to e^+$ conversion is a lepton violating second
 order weak process \cite{KLV94}.
In this definition, the total muon capture 
rate has been written in terms of the well-known Primakoff function
$f_{PR}(A,Z)$ \cite{Pri} which takes into account the effect of the nucleon-nucleon
correlations on the total muon capture rate.
$\vert {\cal M}_{i\to f} \vert ^2$ denotes the square of the partial
transition nuclear matrix element between an initial $|i\rangle$ and a final 
$|f\rangle$ state. This can be written as 
\begin{equation}
\label{10}
 |{\cal M}_{i \rightarrow f}|^2 =\frac{1}{2J_i+1}\sum_{M_f M_i}
| \langle J_f M_f | \Omega | J_i M_i \rangle |^2 
\end{equation}
In our case $|i\rangle=|g.s. \rangle$, i.e.
the ground state of the initial nucleus. The summation
in Eq. (\ref{9}) runs over all states of the final nucleus 
lying up to $\approx 25$ MeV.
($\vert f\rangle \equiv \vert J^\pi _\rho \rangle$. We consider the final 
nuclear states as having well-defined spin (J) and parity ($\pi$). The index 
$\rho$ counts the multipole states).

For the Fermi and Gamow-Teller contributions, which are expected to be the most
important contributions, the square of the matrix element 
$\vert {\cal M}_{i \to f} \vert ^2$ is written as 
\begin{equation}
\label{31a}
|{\cal M}_{i \rightarrow f}|^2= 
\frac{1}{2J_i+1}\, 
\left | \left(\frac{f_V}{f_A}\right)^2
 \langle  f||\Omega_F|| i(gs) \rangle - 
 \langle  f||\Omega_{GT}|| i(gs) \rangle \right |^2
\end{equation}
where $f_V$ and $f_A$ are the usual vector and axial vector coupling 
constants $(f_A/f_V=1.25)$.
By combining Eqs. (\ref{9}) and (\ref{31a}) we see that, for the evaluation of 
the branching ratio $R_{\mu e^+}$, we have to calculate
the reduced matrix elements $\langle f||\Omega_F||i\rangle$ and
$ \langle f||\Omega_{GT}|| i \rangle$ for $|i\rangle=|g.s.\rangle$ 
and $|f\rangle$ any accessible state of the final nucleus.
In the present work these states have been constructed
(for $^{27}$Al and $^{27}$Na systems) in the framework of
the shell model as is described in the next section.

\section{The shell model nuclear wave functions}

The evaluation of the reduced matrix elements
 $\langle f||\Omega_F||i \rangle$
and $\langle f||\Omega_{GT}||i \rangle$ for the $\mu^-\to e^+$
 conversion requires reliable nuclear wave functions. These were  obtained
in the framework of the $1s-0d$ shell model
using the realistic effective interaction of 
Wildenthal \cite{Wil84}. As we have already mentioned, this interaction
 has been tested over many years. It is known to accurately reproduce many
nuclear observables for s-d shell nuclei. The Wildenthal two-body matrix 
elements as well as the single 
particle energies  are determined by least square fits to
experimental data in the region of the periodic table with $A=17-39$. 

The eigenstates of the daughter nucleus $^{27}$Na 
were evaluated in the isospin representation.
For each spin $J_f$ with $T=5/2$ the first states 
reaching up to $E_x=25$ MeV in excitation energy 
were calculated. 
On the other hand, for 
$^{27}$Al we evaluated the ground state $(5/2)^+_{gs}$ with $T=1/2$. We also 
 constructed all the excited (positive parity) states 
up to 5 MeV in order to check our predictions against experiment.
 In Fig. 1 we present the calculated and
measured \cite{NDS} low-energy spectrum of $^{27}$Al up to 5 MeV. 
As can be seen from this figure, the spectrum of $^{27}$Al is well reproduced. 
In the case of the unstable $^{27}$Na isotope the comparison between theory 
and experiment can not be accomplished due to lack of experimental data.

For the special case of the reaction (\ref{Al-react}) studied in the present work,
since $M(^{27}Al)-M(^{27}Na)=-10.6$ MeV, the momentum transfer at which 
our matrix elements must be computed, is given by
\begin{equation}
\label{11a}
p_e=(p_e)_{max}-E_x/c \, , \qquad  (p_e)_{max}=94.5 \, (MeV/c)
\end{equation}

\section{Results and discussion}

As we have mentioned, the primary purpose of the present work is the 
calculation
of the total $\mu^-\to e^+$ reaction rate by summing over partial transition
strengths. As a first step we restricted ourselves in the light neutrino
mechanism. As a result we only dealt with
the Fermi type and the Gamow-Teller type operators
discussed in Sect. 2. In other words we evaluated 
the reduced matrix elements \cite{DIVA00}
\begin{equation}
\label{32a} 
M_F \, = \, \langle f||\Omega_{F} || i(gs) \rangle
\end{equation}
and 
\begin{equation}
\label{33} 
M_{GT}
 \, = \,\langle f||\Omega_{GT} || i(gs) \rangle
\end{equation}
for the transitions between the initial  $|i\rangle=(5/2)^+_{gs}$ and 
all the final  $| f\rangle \equiv |J_\rho^\pi \rangle$ states up to 25 MeV. 
 We evaluated the contributions arising from both the real part and imaginary
 part of the radial function F(r) Eq. (\ref{new-Fr})
which occur in the case of light Majorana neutrinos.

In the calculation of Fermi $|M_F|^2$ and Gamow-Teller  $|M_{GT}|^2$
strengths  we found that the
contribution from  the  multipolarities $L=2$ and $L=4$ is, in general, 
quite small compared to that of $L=0$. This becomes obvious by glancing at
Table 1 where the total Fermi and Gamow-Teller 
strengths with respect to multipolarities $L$ are listed. As can be seen from
Table 1 the Gamow-Teller strength is almost 9 times greater than the Fermi one 
which means that it dominates the total strength.

In order to compare the branching ratio originating from the
$g.s. \rightarrow g.s.$ transition with that associated with
the transition to all final $(5/2)^+$ states, we define, for convenience, the ratio
\begin{equation}
\label{35} 
\lambda \, \equiv \, \frac{R_{gs}}{R}=\frac{ |{\cal M}_{ gs \rightarrow gs}|^2}
{\sum_f  S_{gs \rightarrow f}}, 
\end{equation}
where 
\begin{equation}
\label{35a} 
 S_{gs \rightarrow f} = 
 \left( 1-\frac{E_x }{(p_e)_{max}} \right)^2 |{\cal M}_{gs \rightarrow f}|^2, 
\end{equation}
The  quantity $\lambda$  gives the portion of the $g.s.\rightarrow g.s.$ 
contribution relative to the total
rate here computed by the sum over all partial transitions included in our
model space.
For the $g.s. \rightarrow g.s.$ transition $p_e=94.5$ MeV/c according to Eq.
(\ref{11a}). Since $m_ec <<p_e$ we can consider the approximation
$p_e \approx E_e/c$, which is equivalent to neglecting the electron mass ($m_e$)
in the kinematics of the reaction (\ref{Al-react}).

For the explicit contribution of the excited states to the branching ratio 
we present in Table 2 the sum  
$\sum_f S_{gs \rightarrow f}$ for each set of excited states of given  
$|J_f\rangle$. The second column of Table 2 corresponds to the
contribution from the real part of Eq. (\ref{new-Fr}) while the
third column represents the total contribution coming from the real and imaginary 
parts of Eq. (\ref{new-Fr}). As can be seen from Table 2, the ground 
state transition exhausts a large portion 
(41\%) of the total $\mu^-\to e^+$ reaction rate. 
As can also  be seen from Table 2, 
the main contribution to the total rate comes from
the $(5/2)^+$ multipole states which contribute about 98\% of the total rate.
The rest, $2\%$, originates mainly from 
the $(3/2)^+$, $(7/2)^+$ and $(9/2)^+$    
states.
Another interesting conclusion stemming out of the results of Table 2 is
the fact that, the contribution coming from the imaginary component of the 
radial function F(r) Eq. (\ref{new-Fr}) dominates the total branching ratio. 
In fact the contribution of the real part is about 20 times smaller than 
the corresponding imaginary one.

In order to have an insight about the origin of this difference we studied 
the behavior as a function of the relative coordinate r of the following 
three quantities:

(i) $I_{n n'}(r) = - R_{n 0} R_{n' 0} \ r^2 b \ Im \{F(r)\} $, 

(ii) $S_{n n'}(r) = R_{n 0} R_{n' 0} \ r^2 b \ Re \{F(r)\} $,

\noindent
where $n,n^{'}$ indicate the the nodes in the relative coordinate of the two
nucleon wave function. $Im \{F(r)\}$ and $Re \{F(r)\}$ are the imaginary and
 real parts, 
respectively, of $F(r)$ [see Eq. (\ref{Fr-Re-Im})]. Since there is no 
interference between the real and imaginary parts we find it convenient 
to define $I_{n n'}$ with the (-) sign. For calculations within the $0d-1s$
shell we make the plausible assumption that the main contribution comes from
zero angular momentum states in the relative motion.

(iii) $D_{n n'}(r) = R_{n 0}R_{n' 0} \ r^2 b$, which expresses the nuclear
densities and corresponds to putting $F(r)\approx 1$ (this is the case 
of $0\nu\beta\beta$-decay).

$R_{nl}$ denotes the radial part of the harmonic oscillator wave functions
entering the matrix elements of Eqs. (\ref{32a}) and (\ref{33}). Thus, 
$D_{n n'}$, $I_{n n'}$ and $S_{n n'}$ are dimensionless quantities of
which the squares determine the magnitudes of the two-body matrix elements 
involved in our evaluations of the partial transition rates.
The results obtained for $D_{n n'}(r)$, $I_{n n'}(r)$ and $S_{n n'}(r)$ with 
$n, n' =0,1$, are plotted in Fig. 2 from where we conclude the following:

The variation of $I_{n n'}(r)$ which contains the imaginary part of $F(r)$ 
is in all cases in phase with $D_{n n'}(r)$. The peaks of these 
quantities appear at about the same value of r for each case. On the contrary, 
the variation of $S_{n n'}(r)$ which contain the real part of $F(r)$,
is not always in phase with $D_{n n'}(r)$ and $I_{n n'}(r)$. 
Also the peaks of $S_{n n'}(r)$ do not appear at the same positions of $r$
as those of $D_{n n'}(r)$ and $I_{n n'}(r)$. 
This picture appears in all other cases resulting by using the various
$R_{nl}$ wave functions entering our calculations, even though the
absolute strengths of $D_{n n'}(r)$, $I_{n n'}(r)$ and $S_{n n'}(r)$
are much smaller compared to those of Fig. 2. 

Obviously, the area bounded by the r-axis and $I_{n n'}$ or $S_{n n'}$ gives a 
measure of the contribution of the imaginary or real part of the $\mu^-\to e^-$
operators, respectively, A rough estimation of this area gives for the ratio of
imaginary to real a value of about 4-5 which justifies the factor of about 
15-25 between imaginary and real part contributions to the partial sum 
evaluation of the total $\mu^-\to e^-$ rate.
Of course, this difference of the imaginary part need not apply when the
model is expanded outside our space, i.e. to include $1\hbar\omega$ and 
$2\hbar\omega$ excitations.

At this point we found interesting to study the spreading of the contributions 
through the excitation spectrum of the final nucleus. To this end, in Fig. 3
we plot the distribution of $S_{gs \rightarrow f}$ for all $(5/2)^+$ states.  
Similar pictures are obtained for the other multipole states $(1/2)^+$,
$(3/2)^+$, 
$(7/2)^+$, $(9/2)^+$, $(11/2)^+$, $(13/2)^+$, even though their contribution 
is quite 
smaller than that of $(5/2)^+$. The common feature of these distributions 
is the fact
that for each multipolarity the main contribution comes from low-lying states
and that excitations lying above $\approx$10 MeV contribute negligible 
quantities.
In some cases, as e.g. $(7/2)^+$ the contribution comes from a narrow window
of the excitation spectrum of the final nucleus.

At this stage we should mention that the total rates can also be obtained
in the context of the closure approximation. We remind that according to the
 closure approximation the
contribution of each individual state is effectively taken into account
by assuming a mean excitation energy $\bar{E} = \langle E_f \rangle -
E_{gs}$, and using the completeness relation
$\sum_f|f\rangle\langle f|=1$. Therefore 
$$\sum_f |\langle f |\Omega|i\rangle|^2=\langle i|\Omega^+ \Omega |
i \rangle$$
The matrix element $\langle i|\Omega^+ \Omega |i \rangle$ can be 
written as a sum of two pieces: a two-body term and a four-body one,
that is 
\begin{equation}
\label{cl1}
\langle i|\Omega^+ \Omega |i \rangle= 
\langle i|(\Omega^+ \Omega)_{2b} |i \rangle+
\langle i|(\Omega^+ \Omega)_{4b} |i \rangle
\end{equation}
The total rate evaluation thus involves only the $g.s.$ of the initial nucleus. 
The 4-body piece was neglected in the earlier calculations. Thus, following the
work of Ref. \cite{Eric}, the  relevant two-body operator for the
 process studied in the present work takes the form:
\begin{equation}
\label{cla}
\Omega^+ \Omega =\sum_{i\# j}\tau_- (i) \tau_- (j)
                             \tau_+ (i) \tau_+ (j)
             [\frac{f_V^2}{f_A^2}- \sigma(i). \sigma(j)]
             [\frac{f_V^2}{f_A^2}- \sigma(i) . \sigma(j)]
             (\frac{R_0}{r_{ij}})^2
\end{equation}
 The above equation can be written as follows:
\begin{equation}
\label{clb}
\Omega^+ \Omega =\sum_{i\# j}\tau_-(i) \tau_+(i)
                             \tau_- (j) \tau_+ (j)
 [9+(\frac{f_V^2}{f_A^2})^2- 2(\frac{f_V^2}{f_A^2}+1) \sigma(i) . \sigma(j)]
             (\frac{R_0}{r_{ij}})^2
\end{equation}
To proceed further one makes the simplifying assumption that the matrix element
 is dominated by the spin singlet states. Then the isospin operator counts
 the number of proton pairs in the initial nucleus. The radial part is
 estimated by assuming a uniform nuclear two body density, i.e.
\begin{equation}
\label{clc}
\langle (\frac{R_0}{r_{ij}})^2 \rangle=\frac{1}{(2R_0)^3}\int_{r_c}^{2R_0}
(\frac{R_0}{r})^2 r^2 dr
\end{equation}
where $r_c$ is the hard core radius, assuming for simplicity that the short
range correlations are described by a simple step function. In any case since
$r_c \ll R_0$, the short range correlations can be neglected and to a good
approximation the value of the above integral is $1/4$. We thus find that
\begin{equation}
\label{cld}
\langle i|\Omega^+ \Omega |i \rangle= Z (Z-1)
  (\frac{f_V^2}{f_A^2}+3)^2 \frac{1}{4}
\end{equation}
The above matrix element must be multiplied by the kinematical factor 
$[\langle p_e\rangle/(p_e)_{max}]^2 \approx 0.8$, taking an average excitation energy of about
$20~MeV$. We thus find:
\begin{equation}
\label{cle}
\langle i|\Omega^+ \Omega |i \rangle \approx 0.8 \times 13 \times 12
                                      \times (3.7)^2 \times 0.25=430
\end{equation}
 It is clear, therfore, that the contribution of the excited states found by
 our
state-by-state calculation in the present work  is much smaller than the
predictions found previously by employing closure approximation as
outlined above.

The disagreement appeared between closure approximation and the present
state-by-state calculation can be attributed to the following reasons:

i) The closure approximation takes into account not only the contribution
of $0\hbar \omega$ space but also includes excitations $E\geq 
n\hbar \omega,~~n \ge 1$, as well as the continuum spectrum. A possible extension 
of the $s-d$ model space to include two-particle two-hole states with the
above harmonic oscillator excitations is practically impossible.

ii) In closure approximation the second term in Eq. (\ref{cl1}) which 
includes the four-body forces and which is very complicated, was not taking
into account in the previous calculations. Of course, the obvious question
raised is: how important the contribution of four-body terms is ?

iii) From the ordinary $\mu$-capture it is known that the results of the simple 
closure approximation are quite sensitive to the assumed mean excitation energy 
$\bar{E}$. Since the spectrum of the final nucleus reached by the operators
of $\mu^-\to e^+$ conversion has not yet been observed, $\bar E$ may be naively
estimated from the spectrum of the final nucleus.

Anyway putting our nuclear matrix elements and all the other nuclear
physics input into Eq. (\ref{9}) we get
\begin{equation}
\label{9a}
R_{\mu e^+} = 2.4 ~ \rho | \eta_{eff}|^2 \, . 
\end{equation}
Combining this with the present experimental limit 
$$
R_{\mu e^+} \leq 4.4 \times 10^{-12} \, \, \, \, \cite{Dohmen} \, ,
$$ 
we get
\begin{equation}
\label{9b}
 \rho | \eta_{eff}|^2 \le 1.8 \times 10^{-12} 
\end{equation}
The quantity $ \rho | \eta_{eff}|^2$ depends on the particle model adopted 
as well as the prevailing mechanism for this process and it  will not be further
discussed in this work.

\section{Summary and Conclusions}

In the present work we have investigated the exotic double-charge exchange
neutrinoless muon-to-positron conversion in the presence of nuclei, 
$(A,Z)\mu^-\to e^+(A,Z-2)$.
The appropriate operators have been constructed
considering a variety of gauge model predictions, not only light intermediate
Majorana neutrinos.
We have chosen to study the experimentally interesting 
nucleus $^{27}$Al, since it is going to be used as a stopping target in
the Brookhaven experiment. This nucleus is expected to be described rather
well within the $1s-0d$ shell model, since one 
can use the well-tested Wildenthal $1s-0d$ interaction.

As a first step we restricted to the calculation of the rates in the case of 
light intermediate neutrinos.

>From our results on the reaction $^{27}Al$$(\mu^-,e^+)$$^{27}Na$, we
can conclude the following:

(i) The contribution coming from the Gamow--Teller 
component of the $\mu^-\to e^+$ operator dominates the total rate matrix
 elements.

(ii) The contribution of $L=0$  multipolarity dominates the total rate.

(iii) The total strength, resulting by summing over 
partial transition matrix elements
included in our model space, is much smaller than that found 
previously by using closure approximation. 

(iv) The contribution arising from the imaginary component of the radial
part in the $\mu^-\to e^+$ conversion operator is much larger than the
one of the real component.

(v) The portion of the $g.s.\rightarrow g.s.$ contribution 
(which is proportional to the matrix element $|{\cal M}_{gs \to gs}|^2$)
into the total rate is 41\%. This is good news, since, eventually, the
 experiments will focus on the ground state. 

(vi) By putting our nuclear physics input into Eq. (\ref{9}) we obtain
the limit $\rho | \eta_{eff}|^2 \le 1.8 \times 10^{-12}$. The 
specific limits on $\rho$ and $\eta_{eff}$ depend on the particle model 
assumed and the prevailing mechanism for the $\mu^-\to e^-$ conversion process.

\bigskip

\centerline{\large \bf Appendix }

\bigskip

According to the 
gauge models mentioned in Sect. 2 the transition operator
$\Omega$ can be given in terms of  the following components :

\begin{equation}
\label{A1}
\Omega_{Sa} =\sum_{i\ne j} \tau_-(i) \tau_-(j) e^{\imath{\bf p}_e {\bf r}_i}\,
f(r_{ij}) W_{Sa}(ij),  \quad a=1,2 
\end{equation}
\begin{equation}
\label{A2}
\Omega_{Aa} =\sum_{i\ne j} \tau_-(i) \tau_-(j) e^{\imath {\bf p}_e {\bf r}_i}\,
f(r_{ij}) W_{Aa}(ij),  \quad a=1,2 
\end{equation}
\begin{equation}
\label{jdv1}
\Omega_{T}^{'} =\sum_{i\ne j} \tau_-(i) \tau_-(j) e^{\imath {\bf p}_e {\bf r}_i}\,
f(r_{ij}) W_{S3}(ij) 
\end{equation}
\begin{equation}
\label{jdv22}
\Omega_{A_2}^{'} =\sum_{i\ne j} \tau_-(i) \tau_-(j) e^{\imath {\bf p}_e {\bf r}_i}\,
f(r_{ij}) W_{A3}(ij) 
\end{equation}
where
\begin{equation}
\label{A4}
W_{S1}(ij)=1 
\end{equation}
\begin{equation}
\label{A5}
W_{S2}(ij)={\bf {\sigma}}_i \cdot {\bf \sigma}_j  
=-\sqrt{3}\left [ {\bf \sigma}_i \otimes
{\bf \sigma}_j \right ]^0_0 
\end{equation}
\begin{equation}
\label{A6}
W_{S3}(ij)=3({\bf \sigma}_i\cdot \hat{r}_{ij})({\bf \sigma}_j
\cdot \hat{r}_{ij})-{\bf \sigma}_i
\cdot {\bf \sigma}_j=
\sqrt{6}\left [\sqrt{4 \pi} Y^2(\hat{r}_{ij}) \otimes 
\left [ {\bf \sigma}_i \otimes
{\bf \sigma}_j \right ]^2 \right ]^0_0
\end{equation}
\begin{equation}
\label{A7}
W_{A1}(ij)=\imath {\bf \sigma}_i\times {\bf \sigma}_j=
(-\sqrt{2})\left [ {\bf \sigma}_i \otimes
{\bf \sigma}_j \right ]^1
\end{equation}
\begin{equation}
\label{A8}
W_{A2}(ij)={\bf \sigma}_i- {\bf \sigma}_j 
\end{equation}
\begin{equation}
\label{A9}
W_{A3}(ij)=({\bf \sigma}_i- {\bf \sigma}_j)(\imath\hat{r}_{ij}\times \hat{R}_{ij})
=\sqrt{\frac{2}{3}} 
\left [ \left [    \sqrt{4 \pi}Y^1(\hat{r}_{ij})\otimes \sqrt{4 \pi} Y^1
(\hat{R}_{ij}) \right ]^1\otimes({\bf \sigma}_i- {\bf \sigma}_j) \right ]^0_0
\end{equation}

By Applying  the usual multipole decomposition procedure the
operators  $\Omega_{Sa}$, $\Omega_{Aa}$,  $\Omega_{T}^{'}$ 
and $\Omega_{A_2}^{'}$   read 
\begin{equation}
\label{A10}
\Omega_{Sa}=\sum_{L\Lambda J}   \sqrt{4 \pi} Y^{\Lambda}(\hat{p}_e) \cdot
O_{Sa}^{(L,S)J} \delta_{\Lambda L} \delta_{LJ}
\end{equation}
\begin{eqnarray}
\label{A11}
\Omega_{Aa}=\sum_{LJ\Lambda}\left [ \sqrt{4\pi}
Y^{\Lambda}(\hat{p}_e)\otimes O_{Aa}^{(L,S)J} 
 \right ]^1 \delta_{\Lambda L}
\end{eqnarray}
\begin{equation}
\label{jdv33}
\Omega_{T}^{'}=\sum_{L\Lambda J}   \sqrt{4 \pi} Y^{\Lambda}(\hat{p}_e) \cdot
O_{T}^{(L,S)J}  \delta_{\Lambda J}
\end{equation}
\begin{equation}
\label{jdv44}
\Omega^{'}_{A2}=\sum_{L\Lambda J}   \sqrt{4 \pi} Y^{\Lambda}(\hat{p}_e) \cdot
O_{A2}^{'(L,S)J}  \delta_{\Lambda J}
\end{equation}

The operators $O_{Sa}^{(L,S)J}$, $O_{Aa}^{(L,S)J}$,
$O_{T}^{(L,S)J}$ and $O_{A2}^{'(L,S)J}$ are given 
by the following equations  

\begin{eqnarray}
\label{29d}
O_{S1}^{(L,S)J}
\equiv \Omega_F= 
\sum_{l {\cal L}} A_{l {\cal L} L}
\sum_{i<j}\tau_-(i) \tau_-(j)
f(r_{ij})j_l(\frac{p_er_{ij}}{2})j_{\cal L}(p_eR_{ij}) \nonumber \\
\left [ \sqrt{4 \pi}Y^l(\hat{r}_{ij})\otimes \sqrt{4 \pi} Y^{{\cal L}}(\hat{R}_{ij})
\right ]^J, ~~S=0, J=L
\end{eqnarray}

\begin{eqnarray}
\label{29d1}
O_{S2}^{(L,S)J}
\equiv \Omega_{GT}=
\sum_{l {\cal L}} A_{l {\cal L} L}
\sum_{i<j}\tau_-(i) \tau_-(j)
f(r_{ij})j_l(\frac{p_er_{ij}}{2})j_{\cal L}(p_eR_{ij}) \nonumber \\
\left [ \left [ \sqrt{4 \pi}Y^l(\hat{r}_{ij})\otimes \sqrt{4 \pi} Y^{{\cal L}}(\hat{R}_{ij})
\right ]^L \otimes  (-\sqrt{3})\left [ {\bf \sigma}_i \otimes
{\bf \sigma}_j \right ]^0 \right ]^J,~~S=0, J=L 
\end{eqnarray}

\begin{eqnarray}
\label{jdv3}
O_{A1}^{(L,S)J}=
\sum_{l {\cal L}} B_{l {\cal L} L}
\sum_{i<j}\tau_-(i) \tau_-(j)
f(r_{ij})j_l(\frac{p_er_{ij}}{2})j_{\cal L}(p_eR_{ij}) \nonumber \\
\left [ \left [ \sqrt{4 \pi}Y^{l}(\hat{r}_{ij})\otimes \sqrt{4 \pi} 
Y^{{\cal L}}(\hat{R}_{ij})
\right ]^L \otimes  (-\sqrt{2})\left [ {\bf \sigma}_i \otimes
{\bf \sigma}_j \right ]^1 \right ]^J, ~~S=1, J=L,|L\pm 1|
\end{eqnarray}

\begin{eqnarray}
\label{jdv2}
O_{A2}^{(L,S)J}=
\sum_{l {\cal L}} B_{l {\cal L} L}
\sum_{i<j}\tau_-(i) \tau_-(j)
f(r_{ij})j_l(\frac{p_er_{ij}}{2})j_{\cal L}(p_eR_{ij}) \nonumber \\
\left [ \left [   \sqrt{4 \pi}Y^{l}(\hat{r}_{ij})\otimes \sqrt{4 \pi}
Y^{\cal L}(\hat{R}_{ij}) \right ]^L 
\otimes \left ( {\bf \sigma}_i - {\bf \sigma}_j \right ) \right ]^J,
~~S=1, J=L,|L\pm 1|
\end{eqnarray}

\begin{eqnarray}
\label{30d2}
O_{T}^{(L,S)J}=
  \sum_{l {\cal L}  \lambda} 
  E_{l {\cal L}\lambda L }^{\Lambda} 
  \sum_{i<j}\tau_-(i) \tau_-(j)
  f^{'}(r_{ij})j_l(\frac{p_er_{ij}}{2})j_{\cal L}(p_eR_{ij}) \nonumber \\
 \left [ \left [   \sqrt{4 \pi}Y^{\lambda}(\hat{r}_{ij})\otimes \sqrt{4 \pi}
   Y^{{\cal L}}(\hat{R}_{ij}) \right ]^{L} 
  \otimes \left [ {\bf \sigma}_i \otimes {\bf \sigma}_j \right ]^2 
  \right ]^J~~,\lambda=l,|l\pm 2|,S=2,\Lambda=J
  \end{eqnarray}

\begin{eqnarray}
\label{V3a}
O_{A_2}^{'(L,S)J}=
\sum_{l {\cal L} }\sum_{k \lambda} \sum_{\mu  } 
\Theta_{k \lambda \mu  L}^{\Lambda l {\cal L}}
\sum_{i<j}\tau_-(i) \tau_-(j)
f(r_{ij})j_l(\frac{p_er_{ij}}{2})j_{\cal L}(p_eR_{ij}) \nonumber \\
\left [ \left [   \sqrt{4 \pi}Y^{\lambda}(\hat{r}_{ij})\otimes \sqrt{4 \pi}
Y^{\mu}(\hat{R}_{ij}) \right ]^{L} 
\otimes \left ( {\bf \sigma}_i - {\bf \sigma}_j \right ) \right ]^J,
~~S=1, J=\Lambda
\end{eqnarray}
where

  \begin{eqnarray}
  \label{A.6}
  E_{l {\cal L}\lambda L }^{\Lambda}= 
  D_{l {\cal L}\lambda L }^{\Lambda} A_{l {\cal L} \Lambda}
  \end{eqnarray}
  \begin{eqnarray}
  \label{A.6a}
  D_{l {\cal L}\lambda L }^{\Lambda}=(-1)^{l+{\cal L}+\Lambda}\sqrt{30 \hat{L}
  \hat{\Lambda}\hat{l}}
  \sj{2}{l}{\lambda}{\cal L}{L}{\Lambda}
  \sj{\Lambda}{2}{L}{2}{\Lambda}{0}\langle 20l0|\lambda 0 \rangle
  \end{eqnarray}

\begin{eqnarray}
\label{V4a1}
\Theta_{k \lambda \mu L}^{\Lambda l {\cal L}}= A_{l {\cal L} \Lambda}
I_{k \lambda \mu L}^{\Lambda l {\cal L}} 
\end{eqnarray}
\begin{eqnarray}
\label{V4b}
I_{k \lambda \mu L}^{\Lambda l {\cal L}}= 3\sqrt{2}
(-1)^{l+\lambda+1}\hat{k}
\sqrt{\hat{l}\hat{\cal L}\hat{L}\hat{\Lambda}}
\langle 10l0|\lambda 0\rangle\langle {\cal L}010|\mu0\rangle
\sj{\Lambda}{1}{k}{1}{L}{1}\nonumber \\
\sj{1}{l}{\lambda}{{\cal L}}{k}{\Lambda} 
\sj{\lambda}{{\cal L}}{k}{1}{L}{\mu}
\sj{\Lambda}{1}{L}{1}{\Lambda}{0}
\end{eqnarray}
with

\begin{equation}
\label{v1a} 
A_{l {\cal L} \Lambda}=\sqrt{\frac{\hat{l}\hat{\cal L}}{\hat{\Lambda}}}
\langle l0{\cal L}0|\Lambda 0\rangle (1+(-1)^l)\imath^{l+{\cal L}}
\end{equation}
and
\begin{equation}
\label{v2}
B_{l {\cal L} \Lambda}=
\sqrt{\frac{\hat{l}\hat{\cal L}\hat{J}}{3\hat{\Lambda}}}(-1)^{J+1}
\langle l0{\cal L}0|\Lambda 0\rangle (1-(-1)^l)\imath^{l+{\cal L}}
\end{equation}
($\hat{a}\equiv 2a+1$).

We are now going to discuss the operators entering the leptonic R-L 
interference and in some SUSY mechanisms. Now the relevant operators
may have a time component, which is  small and, in any case, except 
for their radial part is the
same with the $F$ and $GT$ discussed above. They also have a space
component, which is proportional to $\hat{r}_{ij}$. They are vectors, which
yield a scalar, when combined with the leptonic current. They are of 
the form
$$
\sigma_i(\sigma_j\cdot\hat{r}_{ij})+(\sigma_i \cdot \hat{r}_{ij})\sigma_j
-(\sigma_i \cdot \sigma_j)\hat{r}_{ij} \quad \mbox{ and } \quad
i(\sigma_i-\sigma_j)\times \hat{r}_{ij}
$$
The above operators can also be written as
\begin{equation}
\label{vol1}
\sigma_i(\sigma_j\cdot\hat{r}_{ij})+(\sigma_i \cdot \hat{r}_{ij})\sigma_j
-(\sigma_i \cdot \sigma_j)\hat{r}_{ij}=\omega^{'}(0)+\omega^{'}(2)
\end{equation}
and
\begin{equation}
\label{vol2}
i(\sigma_i-\sigma_j)\times \hat{r}_{ij}=\omega^{'}(1)
\end{equation}
where
\begin{equation}
\label{A17b}
  \omega^{'}(k)=\alpha(k)~[ \sqrt{4\pi}Y^1(\hat{r}_{ij}) \otimes T^k(spin)]^1, 
\quad k=0,1,2 \\
\end{equation}
\begin{equation}
\label{A18}
\alpha(0)=-\frac{1}{3\sqrt{3}}~~,~~ T^0(spin)=\sigma_i \cdot \sigma_j
\end{equation}
\begin{equation}
\label{A19}
\alpha(2)=-\frac{2\sqrt{5}}{3}~~,~~ T^2(spin)=[\sigma_i \otimes \sigma_j]^2
\end{equation}
\begin{equation}
\label{A20}
\alpha(1)=\sqrt{\frac{2}{3}}~~,~~ T^1(spin)=\sigma_i-\sigma_j
\end{equation}

The above operators are accompanied by the lepton outgoing waves
\begin{equation}
\label{A21a}
O_{ij}=exp(i{\bf p}_e\cdot {\bf R}_{ij})
[exp(i{\bf p}_e\cdot {\bf r}_{ij})/2+(-1)^{k+1}~ 
exp(-i {\bf p}_e\cdot {\bf r}_{ij})/2]
\end{equation}
The phase of the second term guarantees that the combined operator is 
overall symmetric under the exchange of the particles $i$ and $j$.
The last operator can be brought into the form
\begin{equation}
\label{A21b}
O_{ij}=\sum_{l {\cal L} \lambda \Lambda}
[\beta( l {\cal L} k \Lambda)
j_l(p_e r_{ij}/2)j_{{\cal L}}(p_e R_{ij}) 
[[\sqrt{4 \pi} Y^l(\hat{r}_{ij})
\otimes \sqrt{4 \pi} Y^{{\cal L}}(\hat{R}_{ij})]^ {\Lambda}
\otimes \sqrt{4 \pi} Y^{\Lambda}(\hat{p}_e)]^0
\end{equation}
where
\begin{equation}
\label{A21}
\beta( l {\cal L} k \Lambda)=
[1+(-1)^{l+k+1}]~i^{l+{\cal L}}
              (-1)^{l+{\cal L}}[(2 l +1) (2 {\cal L}+1]^{1/2}
               \langle l0 {\cal L}0|\Lambda 0\rangle 
\end{equation}

Combining the above factors we obtain
\begin{eqnarray}
\label{vol3}
\Omega^{'}(k)=\sum_{LJ\Lambda}\left [ \sqrt{4\pi}
Y^{\Lambda}(\hat{p}_e)\otimes O^{(L,S)J}(k) 
 \right ]^1 
\end{eqnarray}
where
\begin{eqnarray}
\label{A21-app}
O^{(L,S)J}(k)&=& \alpha(k) \sum_{i<j} \tau_-(i) \tau_-(j) f(r_{ij})
\sum_{l {\cal L} \lambda } \beta( l {\cal L} k \Lambda)
\gamma(l,\lambda, {\cal L},L,k,J,\Lambda) j_l(p_e r_{ij}/2) 
\nonumber \\
&& j_{{\cal L}}(p_e R_{ij}) \ [[\sqrt{4 \pi} Y^{\lambda}(\hat{r}_{ij})\otimes 
\sqrt{4 \pi} Y^{{\cal L}}(\hat{R}_{ij})]^L \otimes T^k(spin)]^J
\end{eqnarray}
with
\begin{eqnarray}
\label{A22}
\gamma(l,\lambda, {\cal L},L,k,J,\Lambda)&=& 
    [(3(2 L +1) (2 l +1) (2 J +1)]^{1/2}\\
\nonumber
& &\langle 10l0|\lambda0 \rangle
\sj{1}{k}{1}{J}{\Lambda}{L} \sj{1}{l}{\lambda}{{\cal L}}{L}{\Lambda}
\end{eqnarray}


\clearpage

\begin{table}
\begin{center}
\begin{tabular}{cll}
\hline\hline
       &                        &                       \\
Multipole &\multicolumn{1}{c}{Fermi Contr.}&
\multicolumn{1}{c}{Gamow--Teller Contr.}\\
\hline
       &                        &                       \\
$L=0$  & 8.291                  & 77.272                 \\
$L=2$  & 0.076                  & 0.426                  \\
$L=4$  & 1.553 $\times 10^{-4}$ & 5.515 $\times 10^{-4}$ \\
\hline
\hline
\end{tabular}
\caption{ Fermi and Gamow-Teller transition strengths for various
multipolarities contributing to the partial rate of all the $(5/2)^+$
states. Both the real and imaginary parts of Eq. 
(\protect \ref{new-Fr}) are included.}
\end{center}
\end{table}

\newpage

\begin{table}
\begin{center}
\begin{tabular}{crrr}
\hline\hline
       &                        &                       \\
$\vert f \rangle$ &\multicolumn{1}{r}{Real }
      &Total & $\lambda \ \ (\%)$ \\
\hline
       &             &         &   \\
$gs \to gs$ & 1.38            & 31.59  & 41.44    \\
$1/2$    &3.63$\times$$10^{-3}$ & 0.044  & 0.05    \\
$3/2$    &6.09$\times$$10^{-2}$ &  0.55  & 0.72    \\
$5/2$    &2.24                &  43.52 & 57.08    \\
$7/2$    &3.39$\times$$10^{-2}$ &  0.29   & 0.37    \\
$9/2$    &3.37$\times$$10^{-2}$ &  0.24   & 0.31    \\
$11/2$   &     -             &    -    & -    \\
$13/2$   &     -             &    -    & - \\
         &                   &         & \\
 Total   &  3.75             & 76.23   & 100\\
\hline
          Closure Approx.& &430&100\\
\hline
\hline
\end{tabular}
\caption{Individual contributions  of $\sum_f S_{gs \to f}$
arising from the real and imaginary 
part of the integral $F(r)$ of Eq. (\protect \ref{new-Fr}). As we can see,
the process is dominated by the contribution of the imaginary part.}
\end{center}
\end{table}

\newpage

\clearpage 

\vspace*{0.5cm}
\centerline{\large \bf FIGURE  CAPTIONS}

\vspace{0.5cm}

\noindent
{\bf FIGURE 1} { The calculated (right) and measured (left) \cite{NDS}
energy spectrum  for the lowest positive parity states of $^{27}$Al.}

\vspace{0.5cm}

\noindent
{\bf FIGURE 2.} {Variation of the quantities
$D_{n n'}$, $I_{n n'}$ and $S_{n n'}$ as functions of the relative coordinate
r (see Sect. 5) assuming that the relative zero angular momentum states
dominate. Since the relative motion is indicated in the plots one sees some
 contribution beyond the nuclear radius.
In this figure only the most prominent cases for $n, n' =0,1$ 
are shown. The behavior of $D_{n n'}(r)$, $I_{n n'}(r)$ and $S_{n n'}(r)$ in
the other cases needed for our calculations is similar. }

\vspace{0.5cm}

\noindent
{\bf FIGURE 3.} {Distribution of the transition strength 
 $S_{gs\to f}$ for all $J^\pi = (5/2)^+$ states. Both the real
and imaginary parts of Eq. (\ref{new-Fr}) are included. }

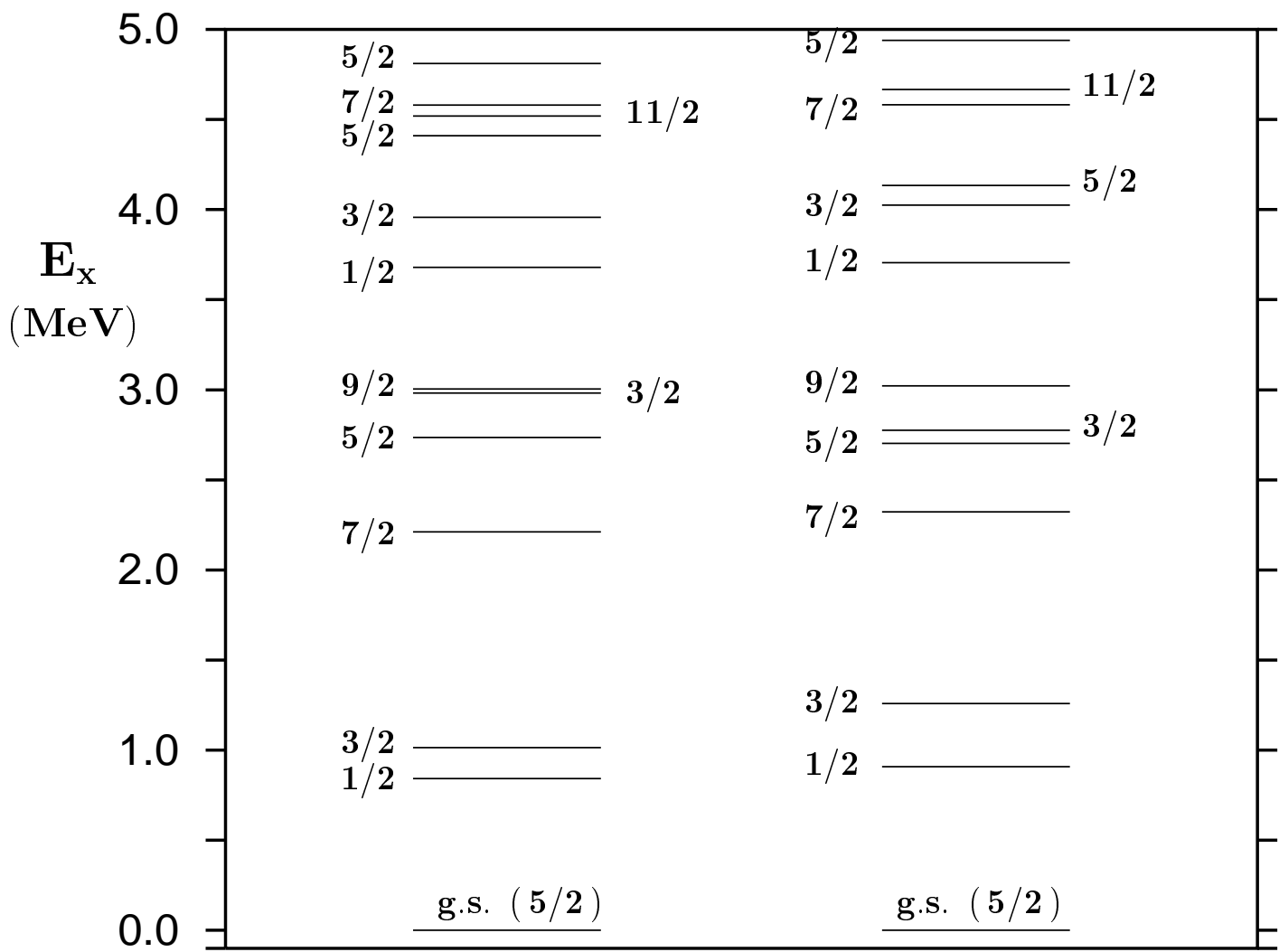
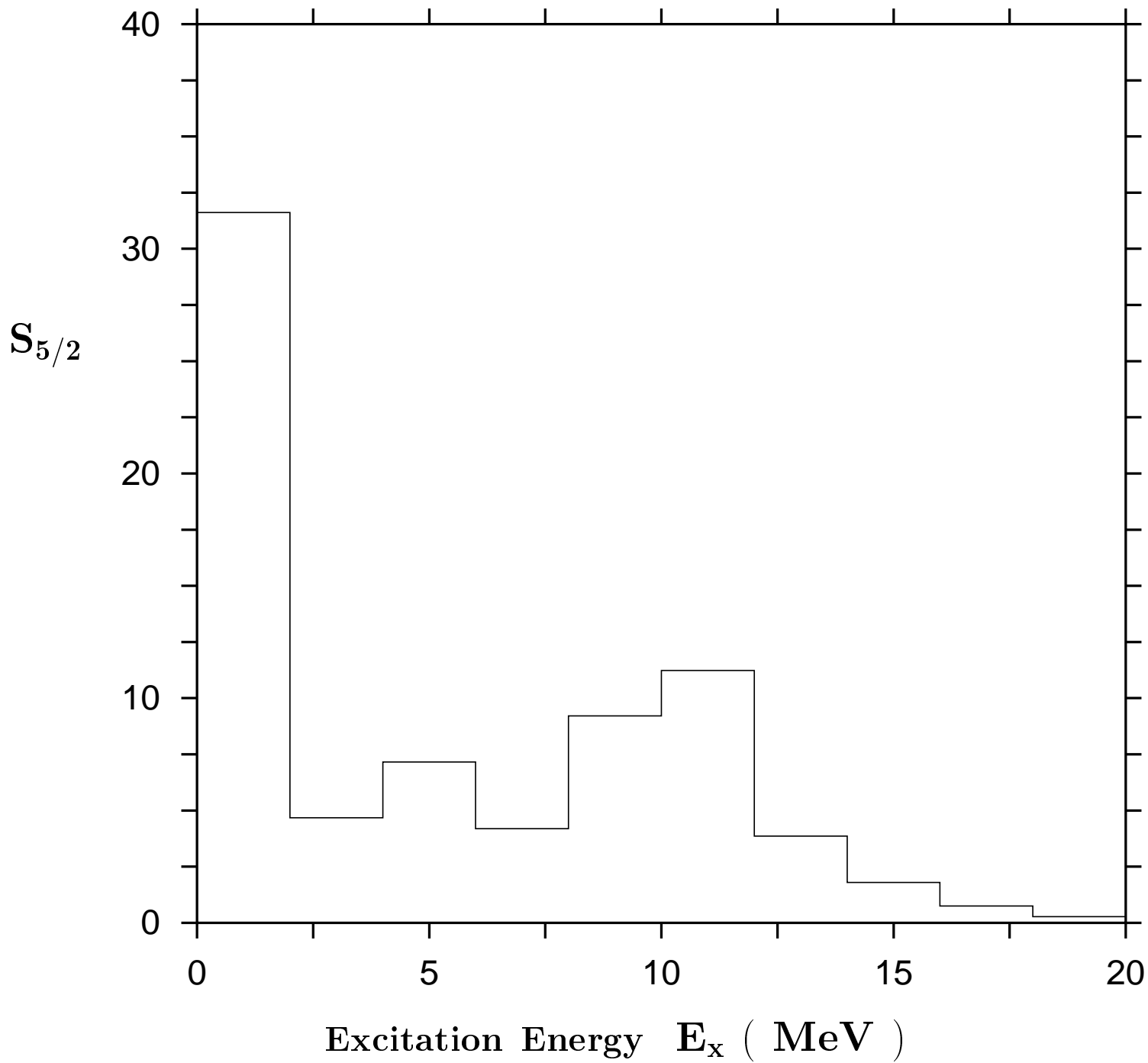
\end{document}